# Title: Recanting witness and natural direct effects: Violations of assumptions or definitions?

Authors:
Ian Shrier MD, PhD

- Ian Shrier: Centre for Clinical Epidemiology, Lady Davis Institute, Jewish General Hospital, and Department of Family Medicine, McGill University, Montreal Canada. (https://orcid.org/0000-0001-9914-3498)

**Corresponding Author:**
Ian Shrier MD, PhD
Centre for Clinical Epidemiology
Lady Davis Institute, Jewish General Hospital, McGill University
3755 Cote Sainte Catherine Road
Montreal, QC  H3T 1E2, Canada
Tel: (514) 340-7563; Email: ian.shrier@mcgill.ca

## Abstract
There have been numerous publications on the advantages and disadvantages of estimating natural (pure) effects compared to controlled effects. One of the main criticisms of natural effects is that it requires an additional assumption for identifiability, namely that the exposure does not cause a confounder of the mediator-outcome relationship. However, every analysis in every study should begin with a research question expressed in ordinary language. Researchers then develop/use mathematical expressions or estimators to best answer these ordinary language questions. When a recanting witness is present, the paper illustrates that there are no violations of assumptions. Rather, using directed acyclic graphs, the typical estimators for natural effects are simply no longer answering any meaningful question. Although some might view this as semantics, the proposed approach illustrates why the more recent methods of path-specific effects and separable effects are more valid and transparent compared to previous methods for decomposition analysis.

## Manuscript

In 2009, Cole and Frangakis discussed whether the "consistency assumption" is an assumption or a definition, and recommended some changes in the way we approach the issue.[1] Pearl called it an axiom, or self-evident truth that must hold to be able to say anything about causality.[2] In this letter, we argue that the additional "recanting witness assumption" to identify the natural direct effect (NDE) and natural indirect effect (NIE) [3] versus the controlled direct effect (CDE) is not an assumption but is actually part of the definition of NDE/NIE. In brief, the "recanting witness assumption" states that the exposure of interest cannot cause a confounder of the mediator-outcome relationship. This "assumption" is required for both Pearl's non-parametric structural equation modeling (NP-SEM) approach and Robins finest fully randomized causal interpretable structural tree graphs approach.[4]

The distinction between definition and assumption is important for four reasons. First, analyses should address important research questions. Just as the target trial approach for observational studies requires "the explicit description and emulation of the target trial",[5] mathematical expressions that do not address the research question and objective as stated in ordinary language are not "definitions" of the research question, and they are likely to provide biased results. Second, we may investigate how results change if we "relax" assumptions", but we do not generally refer to "relaxing definitions". Finally, when we consider the language definition as the starting point, separable effects [6] and path-specific effects[7] are the most transparent decomposition methods and should be the preferred approach where possible.

### *Defining CDE and NDE*

Figure 1A illustrates a causal direct acyclic graph where exposure (A) causes a mediator (M), which causes the outcome (Y). There is a confounder (C) of the M-Y relationship. Compared to the CDE, the "additional assumption" required to identify the NDE/NIE is that A cannot cause C (dotted line).

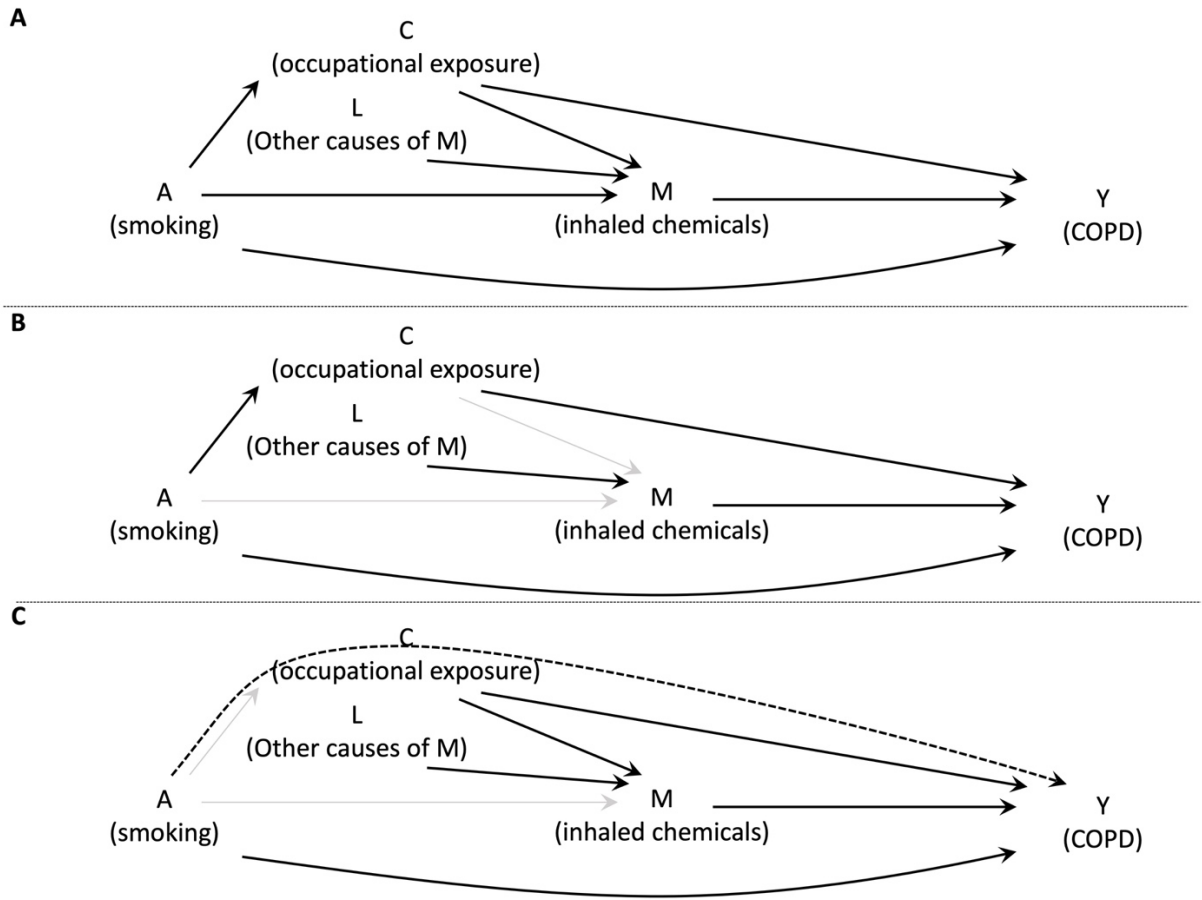

Figure 1: In A, a causal direct acyclic graph where smoking (exposure: A) causes inhaled chemicals (mediator: M), which causes the chronic obstructive pulmonary disease (COPD, outcome: Y). Occupational exposure is a confounder (C) of the M-Y relationship. Compared to the controlled direct effect, the "additional assumption" required to identify the natural direct and indirect effects is that the exposure (smoking) cannot cause the confounder of the mediator-outcome relationship (dotted line from C to M). In B, a complex intervention is applied that blocks (grey arrows) the direct effects of A on M, and the indirect effects of A on M mediated through C by blocking the effects of C on M. This also blocks all effects of C on M even when C is not caused by A. In C, a complex intervention is applied that blocks (grey arrows) the direct effects of A on M, and the indirect effects of A on M mediated through C by blocking the effects of A on C. This also blocks the effects of A on Y that are mediated by C.

The CDE is supposed to measure the effect that would be observed when A is changed from 0 to 1, and M is set to 0 for everyone ($CDE_{M=0}$), or M is set to 1 for everyone ($CDE_{M=1}$).[8] The mathematical expression for this research question is $CDE_{M=0}$ = E[Y | A=1, M=0] – E[Y| A=0, M=0], and $CDE_{M=1}$ = E[Y | A=1, M=1] – E[Y| A=0, M=1].

Pearl defined the NDE as the effect that would be observed when A is changed from 0 to 1, and M is fixed to the value it would have when A=0, even when A=1.[8] Petersen et al. explained that since M takes on the value it would have when A=0, this is equivalent to saying A no longer has a causal effect on M, but M is allowed to vary due to causes other than A.[9] This language definition is particularly important because it relates to developing or improving interventions in the real world. Since we can rarely control for all causes of a mediator, and most biological systems include redundancies, the NDE is usually more clinically relevant than then CDE. [3]

The mathematical expression for the research questions involving NDE is $E[Y \mid A=1, M_{(A=0)}=0]$ – $E[Y \mid A=0, M_{(A=0)}=0]$. This is often referred to as a "cross-world assumption" because in the world where A=1, M is taking on the value from a different world when A=0.[3] It is also referred to as a "recanting witness" problem because C is part of a direct effect (A → C → Y), and an indirect effect (A → C → M → Y).[10] In this sense, C must "recant" its role in the direct effect in order to contribute information to the indirect effect.

Focusing on the language definition, NDE can be rephrased more simply as the effect of changing A from 0 to 1, but A has no effect on M. If (1) A causes C and (2) C causes M, then A causes M. Any identification strategy that only blocks the direct effect of A on M, such as the traditional estimators for NDE/NIE, is not addressing the language definition of NDE/NIE. Thus, the traditional mathematical expression is inconsistent with the language definition and the research question.

When one takes a path-specific effect approach, there are two options to block all the effects of A on M.
1. Figure 1B: We could block the direct effect of A on M (grey arrow), and the direct effect of C on M (grey arrow). This would block all paths through which A causes M. However, the language definition requires that we leave all other mechanisms affecting M intact. If we block the effects of C on M, we are also removing the effects of C on M when C is not caused by A.
2. Figure 1C: We could block the direct effect of A on M (grey arrow), and the direct effect of A on C (grey arrow). Although we are blocking the effects of A on M, we are now also blocking the effects of A on Y that occur through C even though they are independent of M (dashed line).

Therefore, when the exposure causes a confounder of the mediator-outcome relationship, there are no series of causal paths that satisfy the language definition for our research question. The problem occurs because one is trying to apply a mathematical expression derived from a different context (data generating process) that is inconsistent with the research question. As an analogy, parallel lines are sometimes defined as lines that are perpendicular to a single line, and therefore never meet. However, this definition is based on a 2-dimensional model. In a 3-dimensional model such as the earth, two lines (e.g. two longitudes) that are perpendicular to a single line (e.g. 0 degrees latitude) will meet at the north and south pole. The definition that is developed in one context may no longer be applicable if the context changes.

Ensuring a mathematical expression (estimator) reflects the research question is qualitatively different from other assumptions that are made (and can be relaxed) during quantitative analyses. This also highlights an under-recognized value of causal DAGs. Any mediation analysis is really about effects that occur through particular causal paths that involve particular mediators. Both separable effects[6] and path-specific effects [7] are easily illustrated with DAGs, and make the conversion from language definitions to mathematical expressions much more transparent. In brief:
1. We start with ordinary language describing our research question / objective.
2. We trace the path(s) we are interested in capturing.
3. We develop theoretical interventions that remove specific causal pathways (i.e. not of interest) from the exposure to outcome.
4. We ensure our interventions do not affect paths that are of interest.
5. We develop mathematical expressions to estimate the causal effects of the paths of interest.
6. Back-translation into language maximizes the probability that our mathematical expression will match our original objective.